\documentstyle[aps]{revtex}
\input psbox

\begin{document}
\draft
\title{Transport in a Nearly Periodic Potential with a Magnetic Field} 
\author{U. Eckern and K. Ziegler} 
\address{Institut f\"ur Physik, Universit\"at Augsburg, 86135 Augsburg, 
Germany} 
\date{\today} 
\maketitle  
\begin{abstract} 
A two-dimensional gas of non-interacting quasiparticles in a nearly periodic 
potential and a perpendicular magnetic field is studied. The potential is a 
superposition of a periodic potential, induced e.g. by a charge density 
wave or a vortex lattice of a type--II superconductor, and a weak random 
potential due to impurities. Approximating this model by Dirac 
fermions with random mass, random energy, and random vector potential, we 
evaluate the density of states and the Hall conductivity using a 
self-consistent approximation. We obtain a singular average density of 
states $\rho(E)\approx\rho_0+|E|^\alpha$, where $\alpha$ decreases with the 
strength of randomness. The Hall conductivity has a plateau which is 
destroyed for strong tunneling through the saddle points of the nearly 
periodic potential.  
\end{abstract} 
\section{Introduction} 
A two-dimensional system of non-interacting quasiparticles is considered 
in a periodic potential with a strong perpendicular magnetic field. 
The periodic potential can be understood as a pinned charge density wave, 
formed in a two-dimensional low-density electron gas due to Coulomb 
interaction. This model is related to a possible formation of a Wigner 
crystal in a two-dimensional electron gas with a perpendicular magnetic  
field, a problem which has attracted considerable attention recently
\cite{chitra,fertig}. 
Another possible realization of the model is a system of quasiparticles in a 
layered type--II superconductor in the presence of a magnetic field, where
a periodic structure due to the vortex lattice in the Abrikosov phase arises.
Due to impurities, the periodicity of the potential is not perfect in a real
system, and we have to 
incorporate weak random fluctuations of the potential. Therefore, the total 
potential is a superposition of the periodic potential and weak randomness, 
creating a nearly periodic potential.  
 
We start with a continuum model which we map to the network model, following 
the ideas of Chalker and Coddington \cite{chalker/coddington} for the 
quantum Hall system. The resulting model is defined by an evolution  
operator for the quasiparticles on a two-dimensional lattice, where the 
latter is an approximation of the nearly periodic potential. Two 
types of scatterings are possible. 
One describes the scattering along the equipotential lines, while the other  
corresponds to the scattering through the saddle points of the potential. 
This model undergoes a transition between Hall plateaux 
of the integer quantum Hall effect, the quantum Hall transition (QHT), 
when the scattering through the saddle points exceeds a critical value. The 
network model has been very successful as a starting point for numerical 
studies of the QHT \cite{chalker/coddington,lee,cho,hansen}. 
Unfortunately, the model is not easily accessible to analytic methods 
due to its internal lattice structure. However, a large scale 
approximation, for weak disorder, sufficient for the investigation of  
transport properties, has 
been applied to the network model by Ho and Chalker \cite{ho/chalker}. They 
found that the network model is equivalent to a model of Dirac fermions.
This result indicates that the network model is similar to the 
tight-binding model with half flux quantum per plaquette, where the large 
scale approximation also leads to Dirac fermions 
\cite{fisher85,ludwig,zie94}. However, the corresponding randomness in the 
resulting Dirac fermions is different: the tight-binding model with random 
potential leads to Dirac fermions with a random mass whereas the network 
model of the nearly periodic potential leads to Dirac fermions with a 
random vector potential and a random energy. 
 
Dirac fermions with randomness can be treated by field theoretical methods 
\cite{ludwig,mudry,zie972}. A particular case is the purely random 
vector potential, which can be solved exactly by bosonization 
\cite{ludwig,tsvelik}. Other types of randomness, to the best of our
knowledge, cannot be 
treated in this way. Therefore, we will apply a self-consistent 
approximation scheme. In order to control our approximate calculation 
we will compare it with the exact result of Refs. \cite{ludwig,tsvelik}.  
 
The paper is organized as follows. In Sect. II we introduce the continuum 
model, and derive a network model of quasiparticle loops, which is equivalent
to the network model of Chalker and Coddington. In Sect. II.A we briefly
recall a tight-binding approximation of the periodic potential
\cite{ludwig,zie971}. The large scale approximation of the network  
by Dirac fermions is discussed in Sect. II.B. As physical quantities,
we define the average DOS and the Hall conductivity 
$\sigma_{xy}$ in terms of the Green's function. The main part of our 
calculation 
is presented in Sect. III, where a self-consistent approximation for the 
average Green's function is worked out, as well as its consequences for the 
average DOS and the Hall conductivity. Finally, we discuss our results in 
Sect. IV.   
 
\section{The Model and Related Physical Quantities} 
 
We consider the dynamics of a particle with charge $e$ in 
the potential $V(x,y)=V_0\cos(\pi ax)\cos(\pi ay)+\delta V(x,y)$, 
where $\delta V(x,y)$ represents a weakly fluctuating disorder potential. 
We start with a discussion of the classical motion, taking into account the 
Lorentz force created by the perpendicular field ${\bf B}=(0,0,B)$: 
\begin{equation} 
M{d^2 {\bf r}\over dt^2}+{e\over c}{\bf B}\times {d{\bf r}\over dt} 
+\nabla V=0, 
\end{equation} 
where ${\bf r}=(x,y,0)$ is the coordinate vector. 
For a sufficiently small mass $M$ 
the first term can be neglected, and the resulting first order 
differential equations show that the particle follows the equipotential lines 
of the potential: 
\begin{equation} 
{d\over dt}\pmatrix{ 
x\cr 
y\cr 
}={c\over eB}\pmatrix{ 
\partial V/\partial y\cr 
-\partial V/\partial x\cr 
}. 
\end{equation} 
The approximation by the first order differential equation is equivalent 
with neglecting the cyclotron motion of the particle and reducing the 
dynamics to the motion of the guiding center. 
For $B>0$ a particle with energy $E\ne0$ follows closed counterclockwise 
orbits around a hill ($E>0$) or clockwise orbits around a well ($E<0$) 
(Fig. 1). Only a particle with energy $E=0$ can travel 
through the whole two-dimensional space because there are extended 
equipotential lines at $x=(2k+1)/2a$ or at $y=(2k+1)/2a$ ($k=0,1,...$). 
The behavior near $E=0$ can be quantized 
using a lattice approximation. This approximation is based on the orbits of 
quasiparticles at a given energy $E$ which form a network of loops. 
Quantum effects are described by tunneling between these loops in regions 
where the loops are close to each other, i.e. at the saddle points of 
the potential. For $E<0$ the tunneling occurs between the loops in Fig. 2 
whose centers (minima of the potential) are located at $(x,y)$ and  
$(x\pm1,y\pm1)$.  
For $E>0$ we have a complementary situation in which the preferred loops 
appear with the opposite current direction. Apart from this difference, 
the subsequent analysis can be applied to this case as well. We 
will consider $E<0$ in the following. 
 
 
\subsection{Lattice Approximations} 
 
The tunneling can be expressed by discretized scattering 
processes of effective quasiparticle positions on the centers of the loop 
edges, enumerated by $1,...,4$ in Fig. 2. The scattering is 
characterized by a scattering parameter $\beta$ ($0\le\beta\le\pi/4$), 
determined by $V_0$ and $E$. 
Regarding the loop at $(x,y)$ in Fig. 2, there is a quasiparticle state 
${\bf \Psi}(t)\equiv(\Psi_1(t,x,y),\Psi_3(t,x,y),\Psi_2(t,x,y),\Psi_4(t,x,y))$ 
at time $t$ whose components are associated with the edges $j$ on the loop. 
At the later time $t+1$ this state can be scattered along the loop, or can 
tunnel through one of the four saddle points to the loops at  
$(x\pm1,y\pm1)$. For example, the component $\Psi_1(t,x,y)$ can be 
scattered either to $\Psi_2(t+1,x,y)$ with rate $C=\cos\beta$,  
obtaining a phase $\phi_2$, or can tunnel to $\Psi_4(t+1,x+1,y+1)$ with rate 
$S=\sin\beta$, obtaining a phase $\phi_4$. All possible processes can 
be combined as follows: 
\begin{equation} 
{\bf \Psi}(t+1)= 
\pmatrix{ 
0 & {\bf M}\cr 
{\bf N} & 0\cr 
}{\bf \Psi}(t), 
\end{equation} 
where 
\begin{equation} 
{\bf M}=\pmatrix{ 
Se^{i\phi_1}{\bf t}_-^x{\bf t}_+^y & Ce^{i\phi_1} \cr 
Ce^{i\phi_3} & -Se^{i\phi_3}{\bf t}_+^x{\bf t}_-^y\cr 
},\qquad 
{\bf N}=\pmatrix{ 
Ce^{i\phi_2} & Se^{i\phi_2}{\bf t}_+^x{\bf t}_+^y \cr 
-Se^{i\phi_4}{\bf t}_-^x{\bf t}_-^y & Ce^{i\phi_4} \cr 
}, 
\end{equation} 
and ${\bf t}_+^x$ (${\bf t}_-^y$), for example, 
are lattice shift operators in the positive (negative) $x$ ($y$) direction. 
The phases $\phi_1$,..., $\phi_4$ are associated with the centers of the 
loop edges. The weak random potential $\delta V$ induces weakly 
fluctuating phases with $\langle\phi_j\rangle=0$ for $j=1,2,3$ and 
$\langle\phi_4\rangle=\pi$. This choice ensures the presence of half a flux 
quantum per current loop on average. (Notice that the original network 
model \cite{chalker/coddington} has strong disorder due to a uniform 
distribution of the phases $\phi_j$ on the interval $[0,2\pi)$. It is not 
clear if our model with weak disorder belongs to the same universality 
class. Since strong fluctuations of the phase correspond to strong 
fluctuations of the external magnetic field, the fact that a random 
magnetic field is a relevant perturbation \cite{aronov} indicates that 
both models may be qualitatively different.) 
The random potential $\delta V$ also affects the tunneling between the loops. 
This induces randomness in the lattice shift operators 
${\bf t}^{x,y}_\pm$. The latter are statistically independent for different
nearest neighbor pairs on the lattice. 
We believe that $\beta >\pi/4$ is not realistic for the physical model 
because this 
describes a situation where the quasiparticles prefer to tunnel 
through the saddle points rather than go along the equipotential lines. 
 
The (discrete) time evolution of the quasiparticle state on a loop 
can also be described by the evolution operator ${\bf W}$ as 
${\bf\Psi}(t+2)={\bf W}{\bf\Psi}(t)$ \cite{ho/chalker}, with 
\begin{equation} 
{\bf W}=\pmatrix{ 
0 & {\bf M} \cr 
{\bf N} & 0 \cr 
}^2=\pmatrix{ 
{\bf M}{\bf N} & 0\cr 
0 & {\bf N}{\bf M}\cr 
} 
\end{equation} 
 
Assuming (discrete) translational invariance (i.e. $\delta V=0$ in the 
original model), 
we can diagonalize the scattering matrix by applying a Fourier 
transformation $(x,y)\to(k_1,k_2)$. Thus we find the eigenvalues of ${\bf W}$,
\begin{equation} 
\lambda_{1/2}(k_1,k_2)=2SC\cos k_1\cos k_2 
\pm i\sqrt{1-4S^2C^2\cos^2 k_1\cos^2 k_2}. 
\end{equation} 
The corresponding dispersion relation, $E_{1/2}(k_1,k_2)$, obtained via 
$\lambda_{1/2}=\exp(-iE_{1/2})$, has a gap around $E=0$ 
depending on $\beta$ (see also \cite{ho/chalker}). 
The gap vanishes at $k_1=k_2=0,\pm\pi$ for $\beta=\pi/4$, i.e. for $C=S=1/2$. 
Thus the quantum model reflects the behavior of the classical model, where 
infinite equipotential lines exist only at zero energy. The effect of 
disorder on this behavior will be studied subsequently. 
 
We briefly discuss an alternative lattice approximation of the almost 
periodic potential. It is based on the tight-binding representation of the 
continuous model. For this purpose we regard the minima, $V(x,y)=-V_0$, and 
the maxima, $V(x,y)=V_0$, 
of the periodic potential as a lattice. Quasiparticles can hop between 
neighboring lattice points with hopping rate $t$, i.e. for hopping between a 
maximum and a minimum. Moreover, it is reasonable to take the hopping
between next-nearest neighbors into account also which are either pairs of 
maxima or pairs of minima, and associate a hopping rate $t'$ with this 
process. The magnetic 
field leads again to a Peierls phase factor in each of the hopping terms 
\cite{zie971}. Disorder can be introduced by adding random fluctuations 
$\delta V$ 
to the lattice points. This lattice approximation differs from the 
above network model 
because there are two types of hopping now. The nearest neighbor hopping
between 
a maximum and a minimum has been excluded in the network approximation, since 
tunneling is considered only between states at the same energy. 

 
\subsection{Large Scale Approximation: Dirac Fermions} 
 
We are interested in the properties of the quasiparticles on large scales. 
Therefore, we consider the large scale approximation 
of the evolution operator ${\bf W}$ by expanding the non-local part in a 
Taylor expansion ${\bf t}^x_\pm f({\bf r})=t^x_\pm({\bf r})f({\bf r} 
\pm a{\bf e}_x)\approx t^x_\pm({\bf r})[f({\bf r})\pm 
a\nabla_xf({\bf r})]$; for simplicity we set 
$a=1$ in the following. The evolution operator 
then reads for $\beta=\pi/4+\delta\beta$, with $|\delta\beta|\ll 1$,
\begin{equation} 
{\bf M}{\bf N} \approx {\bf 1}+ \pmatrix{ 
i{\bar\phi}-\nabla_x+iA_x & 2\delta\beta+\nabla_y-iA_y\cr 
-2\delta\beta +\nabla_y-iA_y & i{\bar\phi}+\nabla_x-iA_x\cr 
}+{\rm random}\ {\rm terms}\ {\rm from}\ t_\pm^{x,y},
\label{HMN} 
\end{equation} 
and 
\begin{equation} 
{\bf N}{\bf M}\approx {\bf 1}+ \pmatrix{ 
i{\bar\phi}+\nabla_y-iA_y & -2\delta\beta-\nabla_x+iA_x\cr 
2\delta\beta -\nabla_x+iA_x & i{\bar\phi}-\nabla_y+iA_y\cr 
}+{\rm random}\ {\rm terms}\ {\rm from}\ t_\pm^{x,y}, 
\label{HNM} 
\end{equation} 
with the effective random vector potential components 
$A_x=(\phi_1-\phi_3)/2$, $A_y=(\phi_4-\pi-\phi_2)/2$, and with 
${\bar\phi}=(\phi_1+ \cdots +\phi_4-\pi)/2$. 
We can express the results (\ref{HMN}), (\ref{HNM}) using two-dimensional 
Dirac Hamiltonians $H_{MN}$ and 
$H_{NM}$, defined through ${\bf M}{\bf N}\approx{\bf1}-i H_{MN}$ and 
${\bf N}{\bf M}\approx{\bf1}-i H_{NM}$, where 
\begin{eqnarray} 
H_{MN}=-2\delta\beta\sigma_2-(i\nabla_x+A_x)\sigma_3+(i\nabla_y+A_y)\sigma_1 
+{\tilde H}_{MN} 
\nonumber\\ 
H_{NM}=2\delta\beta\sigma_2+(i\nabla_y+A_y)\sigma_3-(i\nabla_x+A_x)\sigma_1 
+{\tilde H}_{NM} 
\end{eqnarray} 
with Pauli matrices $\sigma_j$. The terms ${\tilde H}_{MN}$,  
${\tilde H}_{NM}$ are 
contributions from the random operators ${\bf t}^{x,y}_\pm$. 
Here $m=2\delta\beta$ appears as a Dirac 
mass. Since $\delta \beta\le0$, the Dirac mass cannot be positive. 
Moreover, it is convenient to rotate the $2\times2$ matrices $H_{MN}$ and 
$H_{NM}$ such that the mass term is in the diagonal. This leads to 
\begin{eqnarray} 
H_{MN}=m\sigma_3-(i\nabla_x+A_x)\sigma_1-(i\nabla_y+A_y)\sigma_2 
+{\tilde H}_{MN} 
\nonumber\\ 
H_{NM}=m\sigma_3+(i\nabla_x+A_x)\sigma_1+(i\nabla_y+A_y)\sigma_2 
+{\tilde H}_{NM}. 
\label{diracham} 
\end{eqnarray} 
The random terms, related to the random phases $\phi_j$ and the
random factors $t^{x,y}_\pm$, are expanded in terms of Pauli matrices
as $\sum_{l=0}^3\sigma_l V_l$,
where $l=1,2$ are contributions to the random vector potential, $l=0$ is 
the contribution to the random energy, and $l=3$ the contribution to the 
random mass. ($\sigma_0$ is the $2\times2$ unit matrix.) 
The corresponding expansion for $H_{NM}$ is given by 
$\sum_{l=0}^3\sigma_l V_l'$. It is assumed that 
$V_l$ and $V_l'$ have a Gaussian distribution with zero mean and 
correlations 
\begin{equation} 
\langle V_{l,r}V_{l',r'}\rangle_V=\langle V'_{l,r}V'_{l',r'}\rangle_{V'} 
=\delta_{rr'}\delta_{ll'} g_l, 
\end{equation} 
where $r,r'$ denotes the lattice points. 
In principle, $V_l$ and $V_l'$ are correlated. However, these correlations 
do not play a role here because the Green's function is block-diagonal with
respect to $V_l$ and $V_l'$. 
 
We emphasize that a non-compact continuous symmetry exists in the case 
$g_0=g_3=0$ (pure random vector potential) and $m=0$, namely 
\begin{equation} 
H_{MN}=[(1+\zeta^2)^{1/2}\sigma_0+\zeta\sigma_3]H_{MN} 
[(1+\zeta^2)^{1/2}\sigma_0+\zeta\sigma_3] 
\label{csymm} 
\end{equation} 
and similarly for $H_{NM}$, with 
$-\infty <\zeta <\infty$. This symmetry indicates that the  
pure random vector potential is qualitativly different from the general case 
with $g_0+g_3>0$, e.g. the symmetry is reduced to a discrete one with  
$\zeta=i$ for averages of physical quantities if $m=E=0$. 
 
The time evolution of the quasiparticles in the network is given by 
iterating the time step operator ${\bf W}$. This requires the 
evaluation of $\langle{\bf W}^{n}\rangle_V$ for large values of $n$, since 
we are interested in the long time behavior. We Fourier transform 
${\bf W}^n$ first, and later 
average with respect to the random potential $V$. In a first step, 
\begin{equation} 
\sum_{n\ge 0} {\bf W}^{n}e^{i n\omega} 
=({\bf 1}-{\bf W}e^{i\omega})^{-1}, 
\end{equation} 
where the frequency $\omega$ has an infinitesimal imaginary part  
$\epsilon>0$: $\omega=E+i\epsilon$. 
Now we return to the Dirac Hamiltonian to determine the large scale behavior. 
For small frequencies we obtain the Green's function of Dirac fermions 
\begin{equation} 
({\bf 1}-{\bf W}e^{i\omega})^{-1}\approx\pmatrix{ 
i H_{MN}-i\omega & 0\cr 
0 & i H_{NM}-i\omega\cr 
}^{-1}\equiv iG(\omega). 
\end{equation}  
The evaluation of $\langle G(\omega)\rangle_V$ in a self-consistent approach 
is described below (Sect. III). 
                      
\subsection{Density of States and Hall Conductivity} 
 
According to standard Green's function theory, the average DOS at a given 
energy $E$ is obtained from the Green's function $G$ as 
\begin{equation} \rho(m,E)=\frac{1}{\pi}\lim_{\epsilon\to0} 
\langle {\rm Im} [G_{11,rr}(E+i\epsilon)+G_{22,rr}(E+i\epsilon)]\rangle_V. 
\end{equation} 
The DOS of a pure system ($g_l=0$) is linear around $E=0$ \cite{ludwig} 
and vanishes at $E=0$ for all values of $m$.\\ 
 
The current density in a Dirac model can be calculated directly from the 
Kubo formula \cite{ludwig} or from the response 
to an external static vector potential $q_y$ \cite{bjo}. The effect of 
$q_y$ is a change of the boundary conditions in the $y$-direction, 
a concept extensively used in numerical investigations of Anderson 
localization to study the existence of delocalized states \cite{thoul}. The 
response to the vector potential gives 
the Hall conductivity $\sigma_{xy}$ in terms of Green's functions 
\cite{ludwig,bjo} 
\begin{eqnarray} 
\sigma_{xy} 
={e^2\over h}{i\over q_y}\int\sum_{r'}\Big( 
{\rm Tr}_2\Big\{\sigma_x(H_{MN}-iz-E)^{-1}_{r,r'}(H_{MN} 
-iz-E - q_y\sigma_y)^{-1}_{r',r}\Big\} 
\nonumber\\ 
+{\rm Tr}_2\Big\{\sigma_x(H_{NM}-iz-E)^{-1}_{r,r'}(H_{NM} 
-iz-E - q_y\sigma_y)^{-1}_{r',r}\Big\}\Big){dz\over2\pi}. 
\label{hallcon} 
\end{eqnarray} 
This gives, in the absence of disorder and with periodic boundary conditions, 
and for $q_y\to0$, this leads to 
\begin{equation} 
\sigma_{xy}={e^2\over h}{\rm sign}(-m)\Theta(|m|-|E|), 
\label{conductivity} 
\end{equation} 
where $\Theta$ is the step function. Thus the Hall conductivity has 
the plateau value $e^2/h$ at $E=0$, since $m\le0$ for $\beta\le\pi/4$, while
it is undefined at $m=0$. This behavior of the transport quantity 
$\sigma_{xy}$ is unphysical, and will be replaced by a continuous behavior
when we take disorder into account: then the DOS is found to vanish at $E=0$
for all $m$, which indicates that the Hall current flows on the boundary. 
Therefore, the result (\ref{conductivity}) depends strongly 
on the choice of the boundary conditions \cite{cho,ho/chalker}; 
conditions other than periodic, can suppress $\sigma_{xy}$.
For the tight-binding model of Sect. II.A, $\sigma_{xy}$ was determined for
$E=0$ \cite{zie971} to be 
\begin{equation} 
\sigma_{xy}={e^2\over h}\Theta({t'}^2-V_0^2). 
\label{conductivity2} 
\end{equation} 
Comparing this with the result (\ref{conductivity}), we see that the network 
approximation is similar to the tight-binding approximation provided 
${t'}^2>V_0^2$. 

For the special case of a random vector potential we can evaluate the
dissipative conductivity and the localization length in a simple way. Using
the identity (which holds for $m=0$)
\begin{equation}
G(-\omega)=-\sigma_3G(\omega)\sigma_3,
\end{equation}
we may express the two-particle Green's function
$\langle G(\omega)G(\omega^*)\rangle$ by the product of one-particle Green's
functions on the {\it same} complex half-plane. As a consequence of this
property, the critical properties of the two-particle Green's
function are identical with those of the one-particle
Green's function. This explains the finding that the critical exponent of
the localization length is identical with the decay exponent of $\langle
G(\omega)\rangle$ in an explicit calculation \cite{ludwig}.
                                            
\section{Self-Consistent Approximation} 
 
The full Green's function $\langle G\rangle_V$ can be related to the 
Green's function $G_0$ in the absence of randomness, 
\begin{equation} 
G_0(\omega)=\pmatrix{ 
\omega -\langle H_{MN}\rangle_V & 0\cr 
0 & \omega -\langle H_{NM}\rangle_V\cr 
}^{-1}, 
\end{equation} 
through Dyson's equation:
 
\begin{equation} 
\langle G\rangle_V=G_0+G_0\Sigma \langle G\rangle_V. 
\label{dyson} 
\end{equation} 
The self-energy $\Sigma$, in lowest order perturbation theory, is given 
by $\Sigma\approx\langle VG_0V\rangle_V$. 
Expanding $\Sigma$ in terms of Pauli matrices, 
$\Sigma=\Sigma_0+\Sigma_1\sigma_1+\Sigma_2\sigma_2+\Sigma_3\sigma_3$ 
and, taking into account $\langle G\rangle_V^{-1}=G_0^{-1}-\Sigma$, we see 
that $\Sigma_1$ and $\Sigma_2$ shift the gradient operators in 
$\langle H_{MN}\rangle_V$ and $\langle H_{NM}\rangle_V$. Since 
the parity is conserved for $G_0$ as well as for $\langle G\rangle_V$, these 
contributions must vanish. The remaining contributions, 
$\Sigma_0\equiv -i\eta$ and $\Sigma_3\equiv -m_s$, can be evaluated in a 
self-consistent approximation, where we use second order in perturbation 
theory, and replace $G_0$ in the expression for $\Sigma$ by 
$(G_0^{-1}-\Sigma)^{-1}$, with the result 
\begin{equation} 
\Sigma=\langle V(G_0^{-1}-\Sigma)^{-1}V\rangle_V. 
\label{sce0} 
\end{equation} 
With $g_{ij}=g_i+g_j$, this self-consistent equation leads to two coupled 
equations, 
\begin{equation} 
m_s=m{(g_{12}-g_{03})I\over1-(g_{12}-g_{03})I} 
\label{sce1} 
\end{equation} 
and 
\begin{equation} 
\eta=(\eta-i\omega)gI 
\label{sce2} 
\end{equation} 
with the integral 
\begin{eqnarray} 
I & = & 2\int [(m+m_s)^2+(\eta-i\omega)^2+k^2]^{-1}{d^2k\over(2\pi)^2} 
\nonumber\\ 
& = & 
{1\over2\pi}\ln\Big(1+{\lambda^2\over (m+m_s)^2+(\eta-i\omega)^2}\Big). 
\label{integral} 
\end{eqnarray} 
We introduced $\lambda$ as an ultra-violet cut-off, and $g=g_0+\cdots +g_3$. 
 
For $\omega=0$, there are two solutions of (\ref{sce2}): $\eta=0$, and 
a second one with $\eta\ne0$ and $I=g^{-1}$. The latter implies with 
(\ref{sce1}) a renormalization of the Dirac mass, 
\begin{equation} 
{\bar m}\equiv m+m_s=m{g\over2g_{03}}, 
\label{sce1a} 
\end{equation} 
and with (\ref{integral}), we find 
\begin{equation} 
\eta^2=e^{-2\pi/g}-m^2(g/2g_{03})^2. 
\label{eta} 
\end{equation} 
The multiplicative renormalization of the Dirac mass in Eq. (\ref{sce1a}) 
diverges as $g_{03}\to0$. This indicates a special behavior for a pure 
random vector potential. Although this case is not 
interesting in terms of the network model, it has been discussed extensively 
in the literature \cite{ludwig,mudry,morita}. 
 
For $m=0$ ($\omega\ne0$) we obtain with Eqs. (\ref{sce2}), (\ref{integral}) 
an equation which determines $\eta$, 
\begin{equation} 
e^{2\pi\eta/g(\eta-i\omega)}=1+{\lambda^2\over(\eta-i\omega)^2}. 
\label{eqeta} 
\end{equation} 
Given $\eta$, we can evaluate the DOS as 
\begin{equation} 
\rho(m=0,E)={1\over g}{\rm Re}(\eta), 
\label{dos} 
\end{equation} 
which for $E=0$ vanishes if $m^2\ge m_c^2$, where 
\begin{equation} 
m_c=\pm {2g_{03}\over g}e^{-\pi/g} 
\label{critm} 
\end{equation} 
where we set $\lambda=1$ for simplicity. 
In this case the DOS as a function of $m$ follows a semi-circle 
(Eq. (\ref{eta})), multiplied by $g/2g_{03}$, with the 
maximum at $m=0$ given by $\rho(m=0,E=0)=e^{-\pi/g}/g$. The radius 
of the semi-circle is $2(g_{03}/g)e^{-\pi/g}$, and thus vanishes for 
$g_{03}\to 0$. This means that the DOS vanishes for a pure random vector 
potential ($g_{03}=0$) for any $m\ne 0$, but jumps to  
$\rho(m=0)=e^{-\pi/g_{12}}/ g_{12}$ 
at $m=0$. This again indicates a singular behavior (instability) for the 
pure random vector potential, at least in terms of our self-consistent 
approximation \cite{remark1}. Such behavior was also found in perturbation 
theory around the (integrable) case (random vector potential only), where 
infinitely many relevant operators appear due to the contribution of 
the random Dirac mass \cite{mudry}. 
 
The density of states $\rho(m=0,E)$ behaves smoothly for $E\ne 0$, as shown in
Fig. 5. It shows an effective power law, $e^{-\pi/g}/g+|E|^\alpha$ 
with a cut-off at large $E$ (depending on $\lambda$), except for very small 
and very large 
energies. As can be seen in Fig. 5, the exponent decreases, starting at 
$\alpha=1$ for $g=0$, with increasing $g$.\\ 
 
 
The Hall conductivity can also be calculated within the self-consistent 
approximation. For this purpose we replace the Green's function in  
Eq. (\ref{hallcon}) by $(G_0^{-1}-\Sigma)^{-1}$. This gives an expression
identical to that found for the random Dirac mass \cite{zie94}, except that 
$m_c$ is replaced by the new critical mass (\ref{critm}): 
\begin{equation} 
\sigma_{xy}\approx {e^2\over h}{\rm sign}(-m)\Big[ 
1-(2/\pi){\rm arctan}(\sqrt{m_c^2/m^2-1})\Theta(m_c^2-m^2)\Big]. 
\end{equation} 
This describes a Hall plateau with a continuous decrease to zero near the 
critical scattering parameter $\beta_c=\pi/4$, which implies that disorder 
changes the 
(unphysical) discontinuity of the pure system (\ref{conductivity}) to a 
more realistic behavior. However, for $g_{03}=0$, $g>0$, the critical mass 
$m_c$ vanishes. Thus $\sigma_{xy}$ is undefined for $m=m_c$, similar to 
the pure case. Consequently, the random vector potential alone is not
sufficient 
to create a continuous behavior of $\sigma_{xy}$. However, since there exist 
exact calculations for the random vector potential \cite{ludwig} and numerical
results \cite{morita}, we wish to compare our self-consistent 
approach with those. In order to avoid 
problems with singularities at $m=0$ we assume $m>0$ and set $m\to0$ only 
at the end. In particular, we have to choose
that solution of Eq. (\ref{sce2}) which vanishes with vanishing frequency 
$\omega$. Going back to the self-consistent equation (\ref{sce1}), we can
express $I$ of Eq. (\ref{integral}) 
by the renormalized mass ${\bar m}$ as $I=(1-1/x)/g_{12}$, where 
$x={\bar m}/m$. Consequently, $\eta$ - compare with Eq. (\ref{sce2}) - is
given by $\eta=i\omega(1-x)$. From Eqs. (\ref{sce1}), (\ref{integral}), we
obtain
\begin{equation} 
x=\Big\{1-{g_{12}\over2\pi}\ln\Big[1 
+{1\over x^2(m^2-\omega^2)}\Big]\Big\}^{-1}. 
\end{equation} 
This implies, with $\omega^2\lesssim m^2$, 
\begin{equation} 
x\sim\Big\{1+{g_{12}\over2\pi}\ln\Big[x^2(m^2-\omega^2)\Big]\Big\}^{-1}. 
\end{equation} 
Assuming weak disorder ($g_{12}\ll1$) we exponentiate the right-hand-side 
to get 
\begin{equation} 
x\sim [x^2(m^2-\omega^2)]^{-g_{12}/2\pi} 
\end{equation} 
and
\begin{equation} 
x\sim (m^2-\omega^2)^{-g_{12}/[2\pi(1+g_{12}/\pi)]}. 
\label{ratio}
\end{equation} 
The decay length $\xi$ of the average Green's function reads in terms of 
the self-consistent approximation (in units of the lattice constant) 
\begin{equation} 
\xi=[{\bar m}^2+(\eta-i\omega)^2]^{-1/2}. 
\end{equation} 
Thus, from (\ref{ratio}) and inserting ${\bar m}$, $\eta=i\omega(1-x)$,
we find 
\begin{equation} 
\xi\sim (m^2-\omega^2)^{-1/2(1+g_{12}/\pi)}. 
\end{equation} 
This agrees with the earlier results\cite{ludwig,tsvelik}, obtained from 
a renormalization group argument and a bosonization approach, respectively.
In the latter it is not only restricted to weak disorder, but holds true for
all values of $g_{12}$. 
 
\section{Discussion and Conclusions} 
 
Starting from a two-dimensional system of non-interacting quasiparticles in 
an almost periodic potential with magnetic field, we have derived an effective 
model for large scale properties. The latter is a model of 2D Dirac fermions 
with random mass, random energy and random vector potential. 
Our derivation is analogous to that of Ho and Chalker for a random potential 
\cite{ho/chalker}.  
This model shows a gap due to the periodic potentia,l which disappears if 
the rates for clockwise and counterclockwise scattering of quasiparticles are 
equal. The gap is proportional to the Dirac mass. 
The average Green's function of the random Dirac model 
has been evaluated in a self-consistent approximation. The latter is 
equivalent to the saddle point (or large $N$) approximation of the model 
\cite{zie2}. It gives a self-energy which consists of a multiplicative  
renormalization of 
the average Dirac mass, $m\to mg/2g_{03}$, and a spontaneous creation of a 
complex self-energy.  
 
In terms of the self-consistent approximation only the 
combinations $g_{03}$ and $g_{12}$ enter the results. This supports the 
conjecture \cite{lee} that the randomness of the scattering rates at the 
saddle points is irrelevant (i.e. there is no qualitative difference between 
$g_3>0$ and $g_3=0$) as long as $g_0>0$. In other 
words, the randomness of the phases is sufficient for the qualitative 
description of the QHT, at least for the DOS and Hall conductivity. 
 
The mass renormalization is only a factor $1/2$ 
in the absence of a random vector potential ($g_{12}=0$), but it grows with 
$g_{12}$, and $\eta$ is a function of the renormalized mass and the (real) 
energy $E$. The average DOS is proportional to the real part of $\eta$. It 
shows a semi-circular behavior with width $m_c$, given in  
Eq. (\ref{critm}). This width vanishes if only a random vector potential is 
present, indicating a qualitatively different situation. As a function of $E$, 
it describes an effective power law, 
unless $E$ is very small. The exponent of the power law $\alpha$ decreases
with disorder strength, starting at $\alpha=1$ for the pure system. 
In the case of very weak disorder ($g\sim0$), we take the 
$g^{\rm th}$ power of Eq. (\ref{eqeta}), and expand the result in 
powers of $g$, with the result 
\begin{equation} 
\eta={Eg\over2\pi i}\ln[1-\lambda^2/(E+i\epsilon)^2]+O(g^2). 
\end{equation} 
This yields with Eq. (\ref{dos}) the well-known linear behavior of the DOS 
for $E^2<\lambda^2$, 
\begin{equation} 
\rho(m=0,E)={|E|\over2}\Theta(\lambda^2-E^2) +O(g). 
\end{equation} 
The overall behavior of the DOS is in very good agreement with a numerical
result for a similar Dirac model \cite{morita}. Furthermore, a non-zero 
DOS near the QHT, where $\sigma_{xy}$ deviates from the Hall plateau value 
$e^2/h$, as found by us, also agrees with numerical observations 
\cite{hatsugai,lee/wang}. 
The Hall conductivity $\sigma_{xy}$ has a plateau in the regime where 
scattering is dominated by states localized on the loops. However, in the 
absence 
of disorder, $\sigma_{xy}$ is not defined if scattering is equally probable 
along the loops and through the saddle points ($\beta=\pi/4$). This problem 
disappears in the presence of disorder, since $\sigma_{xy}$ then changes 
continuously from the plateau value $\sigma_{xy}=e^2/h$ to $\sigma_{xy}=0$ at 
$\beta=\pi/4$. 
 
We conclude that the special 
properties of the pure system are cured by the randomness: the vanishing 
average DOS at $E=0$ is elevated to a non-zero value. Although this value is 
exponentially small, it indicates that a band of states exist where the 
quasiparticle can tunnel with high probability through the saddle points of 
the periodic potential (i.e. near $\beta=\pi/4$). 
The tunneling, which can be understood as quantum percolation, obviously 
destroys the plateau of the Hall conductivity. 
At least for the case of a random Dirac mass, it 
is known that part of this band consists of delocalized states \cite{zie972}, 
leading to a non-zero conductivity $\sigma_{xx}\approx(e^2/h\pi)/(1+g/2\pi)$ 
\cite{zie/jug}. Numerical calculations \cite{chalker/coddington,lee,cho} 
indicate that delocalized states also exist in the network model. 
Moreover, for the random vector potential we found a divergent localization
length when $m^2\to\omega^2$ within our self-consistent approach. All
these results are strong hints that there are delocalized states
for the fully random Dirac Hamiltonian, in the regime of strong tunneling
through the saddle points of the nearly periodic potential.

\begin{figure} 
\begin{center}\mbox{\psbox{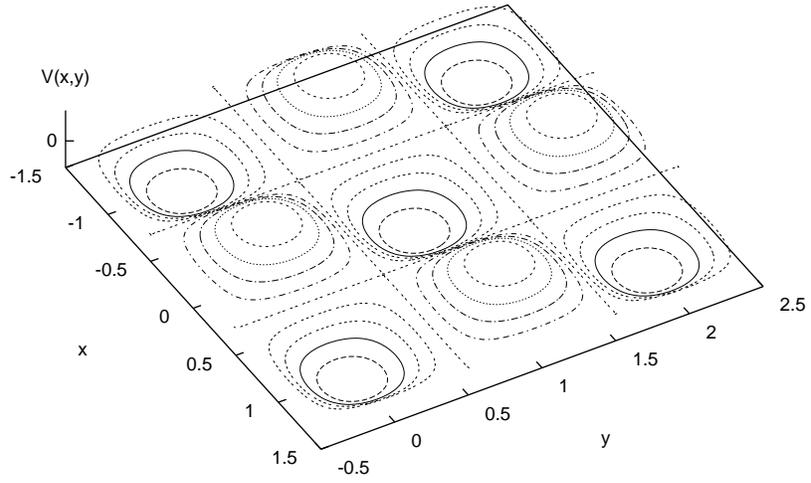}}\end{center} 
\caption{Equipotential lines of $V(x,y)=V_0\cos(\pi x)\cos(\pi y)$ 
(arbitrary units).} 
\end{figure}

\begin{figure} 
\begin{center}\mbox{\psbox{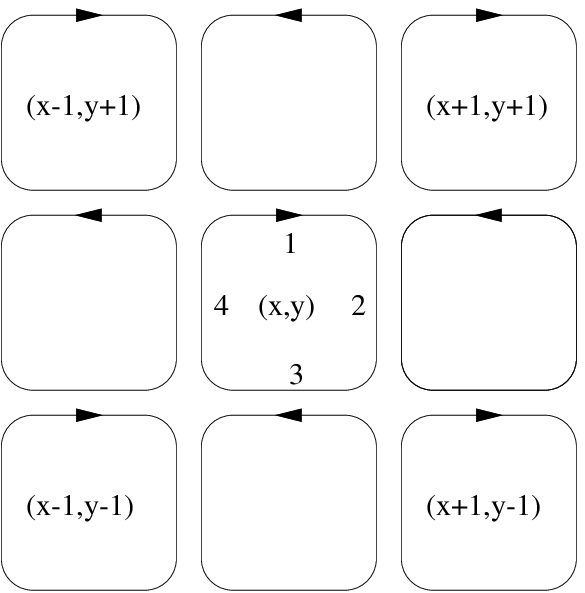}}\end{center} 
\caption{Network of equipotential lines for two fixed energies $\pm E$ close 
to zero. 
The arrows indicate the directions of the quasiparticle motion due to the  
magnetic field $B>0$.} 
\end{figure}

\begin{figure} 
\begin{center}\mbox{\psbox{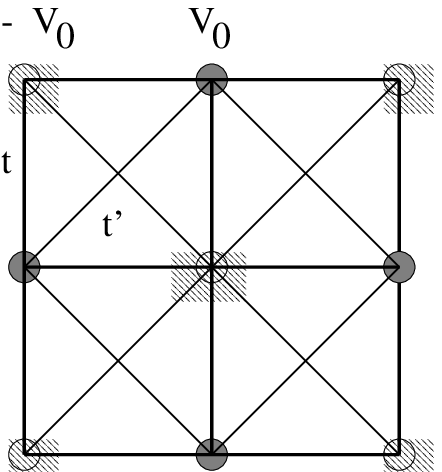}}\end{center} 
\caption{Tight-binding approximation of the periodic potential, where
$t$ is the nearest neighbor and $t'$ the next-nearest neighbor hopping rate.} 
\end{figure}

\begin{figure} 
\begin{center}\mbox{\psbox{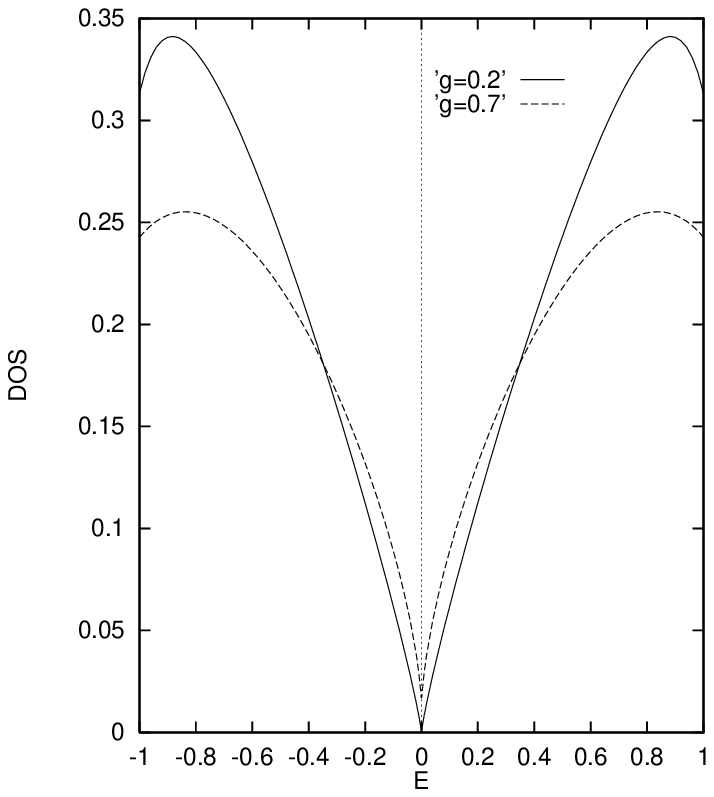}}\end{center} 
\caption{Density of states for weak ($g=0.2$) and intermediate 
($g=0.7$) disorder with scattering parameter $\beta=\pi/4$. The minimum at 
$E=0$ is $e^{-\pi/g}/ g$ (see text).} 
\end{figure}


\begin{references}
\bibitem{remark1} 
The self-consistent approximation is not sufficient in this 
case. An analysis of the corrections in terms of a $1/N$ expansion 
indicates that the singularity 
is related to a Goldstone mode, which is a consequence of the non-compact 
continuous symmetry for the random vector potential (Eq. (\ref{csymm})) 
The combination of the self-consistent DOS at $m=0$ and the logarithmically 
divergent $1/N^2$ correction leads to a power law $\rho\propto |E|^\alpha$.
\end{references}
\end{document}